\documentclass[preprint,12pt]{elsarticle}

\def\lbldef#1#2{\expandafter\gdef\csname #1\endcsname {#2}}

\def\href#1#2{#2}

\usepackage{color}
\usepackage{graphicx}
\usepackage{amssymb}
\journal{Physics of the Dark Universe}

\begin{document}

\begin{frontmatter}

\title{A Cosmological Solution to the Impossibly Early Galaxy Problem}

\author[1]{Manoj K. Yennapureddy\footnote{E-mail: manojy@email.arizona.edu}} 
\author[2]{Fulvio Melia\footnote{John Woodruff Simpson
Fellow. E-mail: fmelia@email.arizona.edu}} 

\address[1]{Department of Physics, The University of Arizona, AZ 85721, USA}
\address[2]{Department of Physics, The Applied Math Program, and Department of Astronomy,
The University of Arizona, AZ 85721, USA}

\begin{abstract}
To understand the formation and evolution of galaxies at redshifts
$0\lesssim z\lesssim 10$, one must invariably introduce specific models
(e.g., for the star formation) in order to fully interpret the data. Unfortunately,
this tends to render the analysis compliant to the theory and its assumptions,
so consensus is still somewhat elusive. Nonetheless, the surprisingly early
appearance of massive galaxies challenges the standard model, and the halo
mass function estimated from galaxy surveys at $z\gtrsim 4$ appears to be
inconsistent with the predictions of $\Lambda$CDM, giving rise to what has
been termed ``The Impossibly Early Galaxy Problem" by some workers in the
field. A simple resolution to this question may not be forthcoming. The
situation with the halos themselves, however, is more straightforward
and, in this paper, we use linear perturbation theory to derive
the halo mass function over the redshift range $0\lesssim z\lesssim 10$
for the $R_{\rm h}=ct$ universe. We use this predicted halo distribution to
demonstrate that both its dependence on mass and its very weak dependence
on redshift are compatible with the data. The difficulties with $\Lambda$CDM
may eventually be overcome with refinements to the underlying theory of
star formation and galaxy evolution within the halos. For now, however,
we demonstrate that the unexpected early formation of structure may 
also simply be due to an incorrect choice of the cosmology, rather than 
to yet unknown astrophysical issues associated with the condensation 
of mass fluctuations and subsequent galaxy formation.
\end{abstract}

\end{frontmatter}

\section{Introduction}
The structures we see today are believed to have grown gravitationally
from tiny fluctuations in the primordial density field. Current theory holds
that perturbations started to collapse once their density exceeded a certain
critical value, forming bound objects that then assembled together with the
surrounding gas and dust to form stars, galaxies and clusters. With dark matter
particles decoupling first from the radiation, the early stages of structure
formation proceeded principally through the condensation of dark matter halos.
Baryonic particles subsequently accreted into the potential valleys created in
this fashion once they themselves decoupled from the relativistic background.

The physics responsible for the formation of galaxies in this scenario is
still not completely understood, but there is general consensus concerning
the rate at which halos formed, specifically their number density distribution
as a function of mass and redshift \cite{Sheth2001,Springel2005,Vogelsberg2014}.
This halo mass function (as it is more commonly known) was first derived
analytically by Press \& Schechter \cite{Press1974} using several simplifying
assumptions, including a spherically symmetric collapse model and a Gaussian
initial density field. But though this analysis predicts a reasonable distribution,
it nonetheless also underpredicts the number of high-mass halos and overpredicts
the low-mass ones compared to detailed numerical simulations. More recently,
Sheth \& Tormen \cite{Sheth2001} have shown that this discrepancy may be mitigated
by adopting an ellipsoidal collapse model rather than spherical. Even so, these
analytical and semi-analytical approaches have for the most part been tested
only against numerical simulations. Unfortunately, while Press-Schechter
underpredicts the number of high-mass halos, Sheth-Tormen apparently overpredicts
them, though a correction factor based on the linear growth rate may have
been found. We shall describe this effect following Equation~(21) below.
It is more difficult to test these semi-analytic approaches using actual 
observations because halos cannot be seen directly. The predicted halo
distribution must be compared to the data indirectly, through the observation
of the galaxy mass function, with an added assumption concerning the evolutionary
relationship between them.

The observed halo formation was recently assessed \cite{Steinhardt2016} using
several previous analyses to compare different techniques for relating the
halo and galaxy distributions. For this purpose, these authors employed high
redshift surveys in the redshift range $z\sim 4-8$, principally The Cosmic Assembly
Near-infrared Deep Extragalactic Survey (CANDELS \cite{Grogin2011,Koekemoer2011})
and the Spitzer Large Area Survey with Hyper-Suprime-Cam (SPLASH \cite{Capak2016}),
to probe the galaxy luminosity and mass functions, from which
the halo distribution may be derived. CANDELS is well suited to find the
lower-mass galaxies because it represents a survey over a small area, whereas
SPLASH has broad sky coverage and can therefore probe the more massive galaxies.

Obtaining the halo distribution and masses from the galaxy distribution presents
quite a challenge. The best way to obtain halo masses from the spatial distribution
of galaxies is via galaxy clustering methods \cite{Hildebrandt2009,Lee2012},
which don't assume any physical properties of the galaxies themselves, though
they must assume a model for the dark matter concentration. Other techniques
use the relationship between the luminosity and stellar masses, obtained
from template fitting \cite{Ilbert2013}. For example the ``abundance matching
technique'' \cite{Finkelstein2015} relates one of the key features in the
luminosity or mass function, such as the knee, to a feature in the halo mass
function, and then matches the galaxy density and dark-matter halo density to
derive halo masses over the whole mass range. Alternatively, one may also assume
that the relations derived at low redshift using luminosity to dark-matter mass
ratios still apply at high redshifts.

But though each of these techniques yields somewhat different outcomes in a quantitative
sense, they all agree qualitatively \cite{Steinhardt2016}. These earlier findings
show quite emphatically that the halo distribution estimated from galaxies at $z\gtrsim 4$
in the CANDELS and SPLASH surveys is inconsistent with the evolution of the halo mass
function and the galaxy luminosity and mass functions predicted by standard
$\Lambda$CDM \cite{Hildebrandt2009,Lee2012,Caputi2015,Finkelstein2015}---a
situation termed ``The Impossibly Early Galaxy
Problem" \cite{Steinhardt2016}. Various possible remedies were considered
by these authors to reconcile the observed and predicted halo mass distributions,
including possible errors introduced in calibrating the data
using relations derived at lower redshifts, which may not be applicable for
$z\gtrsim 4$. None of the remedies worked, however. If anything, this extended
study showed that the high-redshift galaxies appear normal, suggesting that the
relations derived at lower redshifts are probably also applicable at these higher
redshifts.

The tension between the predicitons of $\Lambda$CDM and the `measured' halo
mass function may be resolved with a better understanding of the underlying
physics, e.g., regarding start formation and galaxy evolution. On the other hand,
the current uncertain situation may simply be an indication that there are 
insurmountable problems with the use of $\Lambda$CDM as the background cosmology.
In this paper, we will proceed under this assumption---i.e., that the problems 
elucidated by Steinhardt et al. \cite{Steinhardt2016} are real, and seek to find 
a solution to the surprisingly early formation of massive halos. To balance the 
discussion, however, we acknowledge the fact that this point of view is not 
universally accepted---a situation largely due to uncertainties in the simulations 
used to fully interpret the data and halo mass function. 

Understanding the evolution of galaxies and their
observational signatures, such as their UV luminosity or their redshift-dependent
clustering, necessarily relies on modeling dark-matter evolution \cite{Lacey2011},
and cosmological hydrodynamical simulations \cite{Finlator2011,Waters2016},
complemented by analytical and semi-analytical calculations
\cite{Behroozi2015,Mashian2016}. Complications arise in part because the
observed UV luminosity function depends strongly on redshift (at
least from $z\sim 4$ to $10$), and various combinations of inputs and
assumptions produce degenerate results \cite{Bouwens2015a,Bouwens2015b}.
It is fair to say that the degree of tension between the observations and
predictions of the standard model depends on one's point of view.

But as noted, there are good reasons to suspect that real problems with the formation
of structure do exist in the standard model. Some of these have to do with the
unusually early appearance of supermassive black holes at $z\sim 6-7$
\cite{Melia2013,Melia2015} and galaxies at $z\sim 10-12$ (see refs. cited
in \cite{Melia2014}). In addition, a rather compelling case may be made that
a problem exists \cite{Steinhardt2016} based on the following points:
(1) the halo mass function at
$0 \lesssim z \lesssim 8$ is inferred using 3 or 4 different techniques, not just
one, and all of the results agree at least qualitatively; (2) the fact that these
techniques all require a blending of observational and simulational (i.e.,model-dependent
) factors to arrive at a mutually consistent picture meansre
that it is difficult to understand exactly what the results mean, because such
an approach is very compliant to the assumptions one makes.  For example,
abundance matching forces agreement between observation and theory
even in the absence of a strong physical motivation for the underlying model.
The uncertainties (e.g., in how to match star-forming galaxy UV luminosities
with halo formation in both mass and time) leave unresolved questions
concerning how galactic evolution impacts our understanding of halo
evolution. Nonetheless, forcing consistency between the observations
and predictions of the standard model comes at a considerable cost.

One may understand this situation as follows.
Much of the analysis in this paper is based on the `standard' ratio
of halo to stellar-mass, which arises from two considerations: First
is the expectation that $10\%$ of the baryonic matter condensed into 
stars \cite{Leauthaud2012}. Second, is the ratio of dark matter
to baryonic matter, which is observed to be about 6:1 \cite{Planck2016}. 
As we shall detail below, Steinhardt et al. \cite{Steinhardt2016} 
attempted to reconcile the disparity between theory and observation
by introducing several modifications to the underlying physical processes. 
In order to fit the derived halo mass function in $\Lambda$CDM, however,
they found that only a change by 0.8 dex in the ratio of dark matter to 
baryonic matter would suffice. But such a drastic change could come
about only with a complete absence of dark matter at redshift 8, or 
if essentially $100\%$ of the baryons condensed into stars at higher 
redshifts. Both of these scenarios constitute implausible physics,
such as the need to convert all of the baryons into stars instantly 
upon halo virialization \cite{Steinhardt2016}. Other attempted remedies 
have equally unlikely requirements. So perhaps a better way to characterize
the problem with the halo mass function is to say that it can only be made
consistent with expectations of the standard model with the adoption
of unlikely, new physics.

Given the unsettled debate concerning the formation and evolution
of galaxies, we stress that our focus in this paper is not to model the
galaxies themselves. We merely use some key observations of galactic
profiles to infer the mass and time evolution of halos which, in principle,
constitutes a much simpler, cleaner objective. For a complete assessment
of problems with the formation of structure, it will eventually be necessary
to study both the formation of halos and the galaxies within them, but
this is a much more challenging analysis than we are attempting here.
Such elaborate simulations for the formation and clustering of galaxies
are outside the scope of the present paper. The outcome of this subsequent
work will be reported elsewhere.

In the present context, the difficulty that $\Lambda$CDM has in
accounting for the observed halo mass function has
much in common with the growing tension between the measured cosmological growth
rate, $b\sigma_8(z)$, and its value predicted by the standard model, particularly
in the redshift range $0< z < 1$, where a significant curvature expected in the
functional form of $b\sigma_8(z)$ is absent in the data \cite{Melia2017a}.
Admittedly, the errors in the measured values of
$b\sigma_8(z)$ are still too large to rule out any model, but this is precisely
why a comparison of the measured halo mass function with theory is very probative.
If it turns out that {\it both} the halo distribution at high redshift and
$b\sigma_8(z)$ at lower redshift are in tension with the growth of structure
expected in standard cosmology, a compelling argument can be made that an alternative
expansion scenario must be seriously considered. In this paper, we therefore
compare the measured halo mass function, not only with the prediction of $\Lambda$CDM,
but also with that expected in the alternative Friedmann-Robertson-Walker (FRW) cosmology
known as the $R_{\rm h}=ct$ universe \cite{Melia2007,Melia2016a,Melia2017b,Melia2009,Melia2012}.
It was a direct one-on-one comparison between these two models that highlighted the
greater consistency of the growth rate data with $R_{\rm  h}=ct$ than with
$\Lambda$CDM \cite{Melia2017a}. As we shall see shortly, the measured halo mass
function also provides strong support for $R_{\rm h}=ct$ and, if the tension
between the predictions of $\Lambda$CDM and the data fail to be resolved, may
also eventually argue that $R_{\rm h}=ct$ is favoured over the standard model.

\section{The Halo Mass Function in $\Lambda$CDM}
Let us first see how the halo mass function predicted by the standard model fares
in comparison with the data. Throughout this paper, we use the number density
and halo mass calibrated in ref.~\cite{Steinhardt2016} (and references cited therein)
based on the CANDELS and SPLASH surveys. These data, along with seven theoretical
curves calculated within the redshift range $z=4-10$, are shown in figure~1. The
theoretical curves in this figure are based on the halo mass function estimates
of Sheth \& Tormen \cite{Sheth2001}, using the HMFCalc code developed in
ref.~\cite{Murray2013}. The standard model parameters are assumed to have their
{\it Planck} values, $(h,\Omega_{\rm m},\Omega_\Lambda)=(0.704,0.272,0.728)$
\cite{Planck2016}. Some of these data points were obtained using the clustering
technique and the photometric spectral energy distribution (SED) based on
template fitting. A ratio ${M_H}/{M_*} \sim 70$ \cite{Leauthaud2012}
was used to convert the stellar mass function $M_*$ to a halo mass $M_H$. For
the data points obtained using the UV luminosity function, a conversion
${M_H}/{M_{\odot}} \approx {120L_{UV}}/{L_{\odot}}$ was used to convert
UV luminosity to stellar mass and then to halo mass.

The assumption of a constant ratio $M_H/{M_*}\sim 70$ (and correspondingly
for the UV luminosity) can have a strong influence on the analysis in this
paper, so it is appropriate to question its reliability. The situation today is
somewhat unfortunate in this regard, given that adequate data are lacking to
determine empirically what these ratios are---indeed, whether they are even
constant with redshift. And simulations carried out by several groups do
not appear to produced fully consistent outcomes. We ourselves do not have
a position in this discussion, and eagerly await a resolution in order to
put our calculations on firmer ground. For the time being, our goal is to
find the most reasonable compromise, while maintaining a manageable approach
to avoid tainting our results with excessive (possibly incorrect) detail.

There is some evidence that this ratio is more or less constant based
on fits to the inferred halo to stellar mass ratio at lower redshifts, as 
explained in ref.~\cite{Leauthaud2012}, but its adoption is not universally 
accepted. There has been some debate 
\cite{Conroy2009,Schaerer2010,Lee2012,Stark2013,Speagle2014} regarding whether 
or not these numbers remain unchanged at higher redshifts. Not surprisingly,
there is therefore disagreement between various researchers regarding whether 
or not the halo to stellar mass relation should be parameterized in terms of 
both mass and redshift, rather than just the mass or redshift on its own. For 
example, refs.~\cite{Trac2015,Behroozi2015} used both mass and redshift, 
a conclusion supported by the simulations of Finkelstein et al. 
\cite{Finkelstein2015}, who reported that the halo to stellar mass in their
calculations evolves with both mass and redshift. For example, their results
showed that, at constant UV luminosity, the stellar-to-halo mass ratio increases
with $z$. This trend was also inferred in ref.~\cite{Leauthaud2012}, where 
the halo mass function was parametrized in terms of $z$. But on the other side
of this debate, refs.~\cite{Ma2017,Stefanon2017} reported that the median stellar 
mass to halo mass relation does not evolve strongly in the redshift range
$5\lesssim z\lesssim 12$. Some additional uncertainty was generated by the work 
in ref.~\cite{Mashian2016}, where quite a different approach was used to match 
the observations with theory. These workers abandoned the need for continuous 
evolution altogether, and instead allowed mass to be both added and subtracted 
in order to allow the halo mass function to match the available data at each 
redshift.

There may be an explanation for why there exist two camps in this discussion.
It is certainly true that the ratio of halo mass to luminosity varies in the
simulations of refs.~\cite{Finkelstein2015,Stefanon2017,Ma2017,Ceverino2017,O'Shea2015},
who reported an evolution in both UV luminosity and redshift. Others have pointed
out several caveats, however, mostly having to do with the fact that these conclusions
are based primarily on theoretical modeling. But one cannot always be certain that
all of the necessary physics has been included. Many physical processes appear to
be missing in these simulations, though one does not know for sure how serious any
one of them can be. These include non-equilibrium cooling, photo heating, radiative 
transfer effects, and Pop III star formation, all of which may affect these scaling 
relations.
  
Given this uncertain situation, we will follow the analysis in
ref.~\cite{Steinhardt2016}, whose data we are adopting in this paper,
in which the halo mass was parametrized solely on $z$. This results in 
constant ratios, as are apparently observed at lower redshifts. These 
authors argue that a break down of this condition at higher redshifts,
as suggested by some (perhaps most) simulations, would cause other 
observational signatures to manifest themselves in the evolution of 
growth. But no such effects have been observed, lending some support 
to the assumption of a constant value of these ratios with increasing 
redshift. These authors considered other possible changes to reconcile 
the problem, such as varying the Initial Mass Function (IMF), evolving 
dust corrections, merging and time delays, but to find consistency 
between theory and observations, they were drawn to unphysical conditions, 
some of which we described in the Introduction. They concluded that 
to explain the observed mismatch between the inferred and predicted 
halo mass function, one has to allow for a rapid evolution in the 
IMF or a halo to stellar mass ratio that evolves very rapidly. This 
appears to be unrealistic, given that no such features have been 
verified observationally.

\begin{figure}
\begin{center}
\includegraphics[width=1.0\linewidth]{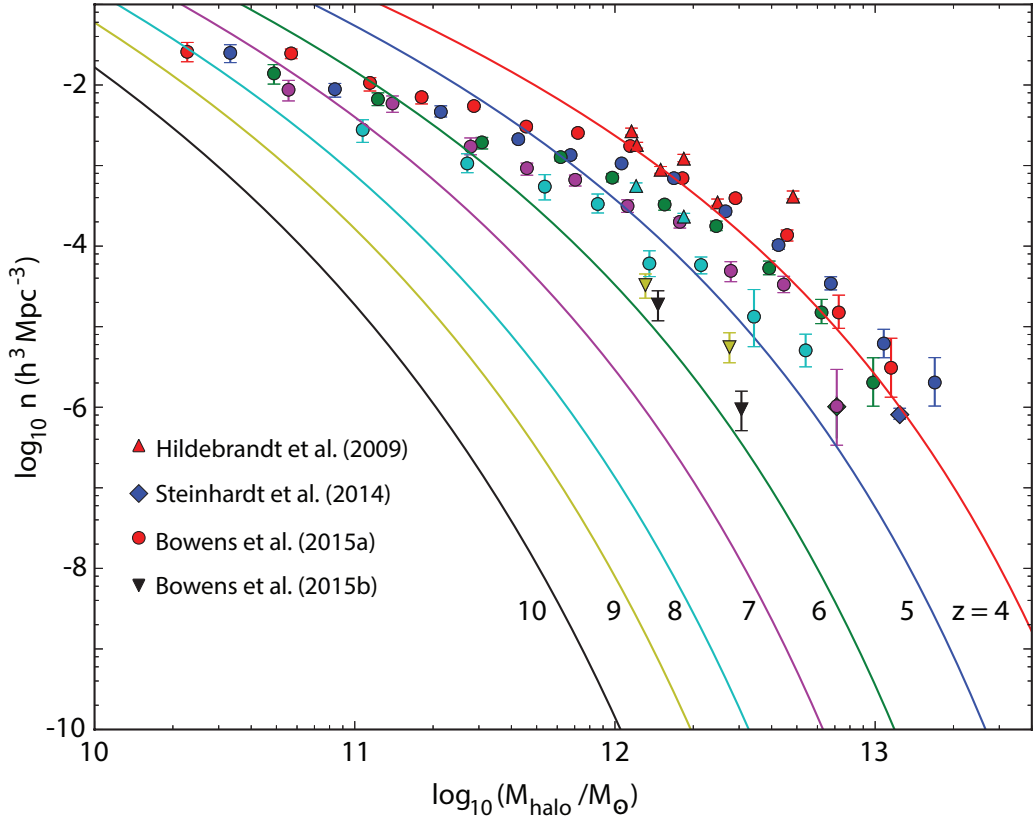}
\end{center}
\caption{Halo mass function inferred from galaxy surveys, as
a function of mass and redshift: $z=4$ (red), $5$ (blue), $6$ (green),
$7$ (magenta), $8$ (cyan), $9$ (yellow), and $10$ (black). Solid curves
represent the theoretical halo number density predicted by $\Lambda$CDM in
this same redshift range, based on the estimates of ref.~\cite{Sheth2001} and
calculated with the HMFCalc code of ref.~\cite{Murray2013}. (Adapted from
ref.~\cite{Steinhardt2016})}
\end{figure}

A quantitative assessment of the degree to which our results
(discussed below) would change with an alternative choice of ratios
shows that an increase in the correlation
${M_H}/{M_{\odot}}\approx {120L_{UV}}/{L_{\odot}}$ by $10\%$
would still permit a reasonable fit (within one order of magnitude)
to the halo mass function at high mass and redshift for
$R_{\rm h}=ct$, but it would not improve the situation for
$\Lambda$CDM. A decrease by $40\%$ of this ratio would still
allow a reasonable fit at high redshift for $\Lambda$CDM,
though not for $R_{\rm h}=ct$. Our results below will show that
the differences in outcome between
the two models is so large that variations in quantities such as
the ${M_H}/{M_*}$ and ${M_H}/{M_{\odot}}\approx {120L_{UV}}/{L_{\odot}}$
ratios are unlikely to be responsible for the full measure of tension
between $\Lambda$CDM and the measured halo mass function.

It is quite evident in figure~1 that the theoretical
predictions of $\Lambda$CDM are inconsistent with the observational data.
The $\Lambda$CDM model predicts a sharp evolution in the density of massive
halos beyond redshift 4, but as one can see from this figure, the number density
of halos is far greater than predicted at these high redshifts. 
We shall see shortly that, whereas $\Lambda$CDM is left without any
viable explanation for the measured growth rate and halo mass function, both of
these problems are resolved in the $R_{\rm h}=ct$ cosmology. In the next section,
we will discuss the growth of perturbations in $R_{\rm h}=ct$ and use this result to
derive the halo mass function assuming the ellipsoidal collapse model of
Sheth \& Tormen \cite{Sheth2001} for the overdense regions.

\section{Linear Perturbation Theory and Halo Mass Function in $R_{\rm h}=ct$}
\subsection{Linear Perturbation Growth}
The initial density field grew as a result of self-gravity, while the Hubble
expansion and pressure effects diluted it. Which of these factors dominated
at any given time determined the evolution of the initial density fluctuations.
To follow the evolution in density, one must solve the relativistic
perturbed equations because the matter may be coupled to relativistic
components, including radiation. Much of the ground work for this theory
has been laid out in ref.~\cite{Melia2017a}, and we here adopt some
of the principal results of that work to trace out the evolution of the
perturbations $\delta \equiv {\delta \rho}/{\rho} \leqslant1$, where the
density and pressure are written
\begin{equation}
\rho=\rho_0+\delta \rho\;,
\end{equation}
\begin{equation}
p=p_0+\delta p\;.
\end{equation}

These fluctuations perturb the metric $g_{\alpha \beta}$ and stress-energy tensor
$T_{\alpha \beta}$, all of which are included in the set of linearized relations
derived from Einstein's equations. Many of the details of this derivation have
appeared previously in
refs.~\cite{Weinberg1972,Landau1975,Peebles1980,Peebles1993,Press1980,Kolb1990,Padmanabhan1993,Coles1995,Peacock1999,Liddle2000,Tsagas2008}.
For example, it is not difficult to show that in the covariant Lagrangian approach,
where the proper time and cosmic time are related via the gauge transformation
\begin{equation}
\frac{d\tau}{dt}=1-\frac{\delta p}{\rho + p}\;,
\end{equation}
one has \cite{Padmanabhan1993,Liddle2000}
\begin{equation}
\frac{d\delta \rho}{dt}=-3H_0\delta \rho-3\delta H(\rho_0+P_0)\;.
\end{equation}
In addition, perturbations to the metric result in
\begin{equation}
\frac{d\delta H}{dt}+2H_0 \delta H+\frac{4\pi G}{3c^2}\rho_0\delta+\frac{v_s^2}{3(1+w)}D^2\delta=0\;,
\end{equation}
where $\delta H$ is the corresponding perturbation to the Hubble constant, i.e.,
$H(t) = H_0(t)+\delta H(t)$, and $w$ is the equation of state parameter, defined
as $p=w\rho$. Also, $v_s^2\equiv \partial p/\partial \rho$ is the sound speed squared.

The $R_{\rm h}=ct$ cosmology satisfies the zero active mass condition, $\rho + 3p=0$,
at all times, so the Universe is dominated by dark energy and (baryonic $+$ dark) matter
at low redshifts ($z \lesssim 10$), and by dark energy and radiation in the early
Universe. Since we are only considering redshifts $z\lesssim 10$ in this paper,
we focus exclusively on the growth of linear perturbations in the matter and dark energy
dominated era. We further assume that dark energy acts as a smooth background, so the
perturbations are strictly due to fluctuations in the matter density. As discussed in
ref.~\cite{Melia2017b}, the zero active mass condition requires us to retain the leading second
order terms as well, which are usually unimportant relative to the first order terms.

The fluctuation growth is characterized by the following expressions:
\begin{equation}
\dot \delta=-H_0\delta-2\delta H-3 \delta H \delta\;,
\end{equation}
\begin{equation}
\ddot \delta+3H_0 \dot \delta-\frac{3}{2}(\dot \delta^2-H_0 \delta \dot \delta+H_0^2\delta^2)=v_s^2D^2\delta\;.
\end{equation}
As is well known, the second term on the left-hand side represents the
dilution due to Hubble expansion, while the third term represents
growth due to the gravitational instability. The right-hand side accounts
for the effects of pressure within the perturbed fluid, sometimes producing
acoustic oscillations. But notice that with the zero active mass condition,
the gravitational growth term is zero to first order. This is the most
significant difference between the $\Lambda$CDM and $R_{\rm h}=ct$ predictions.
While the growth equation in the standard model implies a strong gravitational
instability and, therefore, a strong evolution of the halo mass function with
redshift (see figure~1), $\delta(t)$, and hence the halo growth rate, are
much weaker functions of redshift in $R_{\rm h}=ct$.

We may find a relatively straightforward solution to Equation~(7) by utilizing
the weak dependence of $\delta(t)$ on $t$. If we adopt the ansatz
\begin{equation}
\delta (t)\sim t^\alpha\;,
\end{equation}
with $|\alpha| \ll 1$, then the third term on the left-hand side of Equation~(7) gives
\begin{equation}
\frac{3}{2}(\dot \delta^2-H_0 \delta \dot \delta+H_0^2\delta^2)=\frac{1}{t^2}
(\alpha^2 \delta^2-\alpha\delta^2+\delta^2)\approx B t^{-2}\alpha
\end{equation}
(where $B$ is itself much smaller than $1$), so that
\begin{equation}
\ddot \delta+3H_0 \dot \delta-\frac{B}{t^2} \delta=v_s^2 D^2 \delta\;.
\end{equation}
We may then follow the conventional procedure of solving this equation
using modal analysis with the Fourier decomposition
\begin{equation}
\delta _k(t)=\int \delta(x^\alpha) e^{i\textbf{k.x}}d^3x\;.
\end{equation}
Since, in addition, one has $a(t)={t}/{t_0}$ and $H_0(t)=1/t$ in
$R_{\rm h}=ct$, we find that
\begin{equation}
\frac{d^2 \delta _k}{dt^2}+\frac{3}{t} \frac{d \delta_k}{dt}-
\frac{B}{t^2} \delta_k=-\frac{k^2}{a^2}v_s^2\delta_k\;.
\end{equation}
In the redshift range $z\lesssim 10$, the Universe is dominated by
matter and dark energy, so one may ignore the contribution of radiation
and write $\rho\approx\rho_{\rm m}+\rho_{\rm de}$. Also, since the
perturbation is assumed to contain only matter, $v_s \approx 0$, and therefore
\begin{equation}
\ddot \delta_k +\frac{3}{t}\dot \delta_k-\frac{B}{t^2}\delta_k=0\;.
\end{equation}
This equation has a polynomial solution,
\begin{equation}
\delta_k(t) =(C_1t^{-2}+C_2t^{B/2})
\end{equation}
where $C_1$ and $C_2$ are constants that depend on the initial conditions.
The first term describes a decaying mode, whereas the second term is the
growing---actually, quasi-steady---mode. Ignoring the decaying mode, we have
\begin{equation}
\delta_k(t)\approx \delta_k(t_0)\bigg(\frac{t}{t_0}\bigg)^{B/2}
\end{equation}
or, equivalently,
\begin{equation}
\delta_k(z)=\delta_k(0)(1+z)^{-B/2}\;.
\end{equation}
Using a spherical top-hat window function $W_R(x)$ to filter the fluctuations
$\delta(x)$ on a scale of radius $R$, the variance of the fluctuations is given as
\begin{equation}
\sigma_R^2(R,z)\equiv \frac{b^2(z)}{2\pi^2}\int_{0}^{\infty} k^2 P(k)W^2(k,R)dk\;,
\end{equation}
where $b(z)$ is the usual growth factor which, from Equation~(16), may be written as
\begin{equation}
b(z)=(1+z)^{-B/2}\;.
\end{equation}
The top-hat filter in Fourier-space is given by
\begin{equation}
W(k,R)=\frac{3\Big[ \sin(kR)-kR\cos(kR)\Big]}{(kR)^3}\;.
\end{equation}
Since one can readily see from Equation~(9) that $|B|\ll 1$, it is clear that
the fluctuations grow very slowly at low redshifts. This is the principal
feature that makes the predictions of $R_{\rm h}=ct$ consistent with the
$b\sigma_8$ data at $z\lesssim 1$, while $\Lambda$CDM predicts a significant
curvature in this function that is not confirmed by the observations.

\subsection{The Halo Mass Function}
One can address questions concerning the fraction of matter bound in
structures and their distribution using the halo mass function, first
derived by Press and Schechter \cite{Press1974} assuming spherical collapse and
a Gaussian initial density field. It is well known, however, that the Press-Schechter
mass function overpredicts the number of high-mass halos and underpredicts the number
of low-mass ones. This problem can be resolved using an ellipsoidal collapse model,
rather than spherical, producing the Sheth-Tormen mass function \cite{Sheth2001},
given as
\begin{equation}
f(\sigma)= A\sqrt{\frac{2a}{\pi}}\bigg[1+\bigg(\frac{\sigma^2}{a\delta_c^2}\bigg)^p
\bigg]\frac{\delta_c}{\sigma}\exp\bigg[-\frac{a\delta_c^2}{2\sigma^2}\bigg]\;,
\end{equation}
where $A=0.3222$ is a normalization constant, $a=0.707$ and $p=0.3$.
This mass function is related to the number density of halos with masses
less than $M$ as follows:
\begin{equation}
f(\sigma)=\frac{M}{\rho_0(z)}\frac{dn(M,z)}{d\ln \sigma^{-1}}\;,
\end{equation}
where $\sigma$ is the variance of the fluctuations in Equation~(17).

A quantitative assessment of the merits of Press-Schecter
versus Sheth-Tolman is hard to come by because tests of these 
analytic functions necessarily rely on numerical simulations, 
rather than actual model-independent measurements. However, in 
their Bolshoi simulation completed just a few years ago, Klypin 
et al. \cite{Klypin2011} examined in detail how the Sheth-Tormen
prediction compares with their results as a function of mass and 
redshift. They found that discrepancies were small, i.e., less 
than $\sim 10\%$, at $z\approx 0$ for masses in the range 
$\sim (5\times 10^9-5\times 10^{14})\; M_\odot$. But the Sheth-Tormen
analytic function over-predicts the halo abundance at higher redshifts. 
For example, at $z\sim 6$, Sheth-Tormen over-predicts the number of halos by about
$50\%$ compared to the simulation for masses $\sim 10^{11}-10^{12}\;M_\odot$. 
This prediction worsens by an order of magnitude at $z\sim 10$.

In the context of $\Lambda$CDM, a remedy for this defect is to include 
a multiplicative (correction) factor (introduced in ref.~\cite{Klypin2011}) 
that brings the Sheth-Tormen approximation to $\lesssim 10\%$ deviation 
compared to the simulations for masses $\sim (5\times 10^9-5\times 10^{14})
\; M_\odot$, and redshifts $0\lesssim z\lesssim 10$: 
\begin{equation}
F(b[z])={(5.501 b[z])^4\over 1+(5.500 b[z])^4}\;,
\end{equation}
where $b(z)$ is the growth factor in Equation~(18) (normalized to $1$ 
at $z=0$). But as we have pointed out, $b(z)$ is a very slowly-growing
function of $z$ in the $R_{\rm h}=ct$ universe, so the impact of such
a correction factor is minimal at best. Since an actual
numerical simulation analogous to Bolshoi has not yet been carried out
for $R_{\rm h}=ct$, one does not know yet whether the discrepancies
seen for $\Lambda$CDM persist here. Perhaps the much weaker dependence of
$b(z)$ on $z$ mitigates the over-prediction seen for the standard model.
For the purpose of this paper, given (1) that the results published in
ref.~\cite{Steinhardt2016} (and reproduced in fig.~1) are based on the
Sheth-Tormen approximation without a correction factor, and (2) that
the linear growth rate $b(z)$ changes very little with redshift in
$R_{\rm h}=ct$, thus making the correction factor $F(b)$ virtually
ineffective, we will continue to use the Sheth-Tormen result in 
Equation~(21) without the possibly important modification introduced
in Equation~(22). This issue will be revisited in future work once
a full numerical simulation analogous to Bolshoi will have been
carried out for $R_{\rm h}=ct$.

In terms of the impact of this approach on the respective results
in $\Lambda$CMD and $R_{\rm h}=ct$ are concerned, one can see right
away from figure~1 that reducing the Sheth-Tormen prediction for
the number of halos at high mass only makes things worse compared
to the data shown in this plot, since the standard model's prediction
already falls well below the measurements. For $R_{\rm h}=ct$, on the
other hand, we shall see in figure~2 below that the theoretical prediction
is very good for the lower masses, but misses progressively more and
more towards the high-mass end. The deviation at a halo mass $\sim
10^{13}\;M_\odot$ is a factor of $10-20$, suggestively close to the
over-prediction found in \cite{Klypin2011}. It will be interesting
indeed to see in future work if a correction factor similar to that 
in Equation~(22) brings the results of $R_{\rm h}=ct$ completely in 
line with the data at all masses.

An additionial caveat with our assumptions in this paper is that simulations
produce results that are not always consistent with each other when different `halo
finders' are used to identify the mass condensations. The recently performed Bolshoi 
calculations \cite{Klypin2011} indicate that Sheth-Tormen (used in our analysis)
overpredicts the number of halos found in Bolshoi at $z=8.8$ by a factor of 4-6 
when the Spherical Over (SO) density halo finder algorithm is used, whereas 
Sheth-Tormen is actually consistent with the results based on the Friends of 
Friends (FOF) algorithm. Worse, the inferred number and mass distributions are
inconsistent with each other in the simulation when the results of FOF are
compared with those of SO. Bolshoi found that both algorithms identified 
the same distinct halos, but FOF assigned larger masses to some of them. 
On average, the masses of halos found with FOF was $1.4$ times larger than 
those based on SO for the same halos. Masses obtained using FOF and SO 
tend to be consistent at lower redshifts, but deviate from each other
at higher redshifts. Noting this as a potential problem to address in
future, more elaborate treatments of the analysis we attempt here, 
we also acknowledge the fact that these inconsistencies are still
small compared to other potential systematic issues entering our
investigation, such as the poorly known star formation rate and galaxy 
evolution models used in the `measurement' of the halo mass function. 
We will therefore proceed without attempting to address these issues here.

\section{A Cosmological Resolution of the Impossibly Early Galaxy Problem}
Before comparing the predicted halo mass function with the observations, we must
recalibrate the data to take into account the differences in differential comoving
volume between the two models. This is most easily done using the following
expressions for the comoving distance:
\begin{equation}
D^{\Lambda{\rm CDM}}_{\rm com}=\frac{c}{H_0}\int_{0}^{z}\frac{du}{\sqrt{\Omega_{\rm m}
(1+u)^3+\Omega_{\rm r}(1+u)^4+\Omega_\Lambda}}\;,
\end{equation}
and
\begin{equation}
D^{R_{\rm h}=ct}_{\rm com}=\frac{c}{H_0}\ln(1+z)\;.
\end{equation}
Note that here $H_0$ refers to the Hubble constant today. The conversion factor
to recalibrate the data is then simply

\begin{equation}
\frac{{dV^{\Lambda{\rm CDM}}_{\rm com}}/{dz}}{{dV^{R_{\rm h}=ct}_{\rm com}}/{dz}}=
\left(\frac{D^{\Lambda{\rm CDM}}_{\rm com}}{D^{R_{\rm h}=ct}_{\rm com}}\right)^2 
\frac{{dD^{\Lambda{\rm CDM}}_{\rm com}}/{dz}}{{dD^{R_{\rm h}=ct}_{\rm com}}/{dz}}\;.
\end{equation}
The impact of this recalibration may be gauged via a comparison of the data
plotted in figures~1 and 2. Within the redshift range $0\lesssim z\lesssim 10$,
the differential comoving volumes in $R_{\rm h}=ct$ and $\Lambda$CDM differ
by less than $\sim 10\%$. One sees similarly small shifts in $n$
between these two plots.

A second factor one must take into account while calculating the halo distribution
is the value of $\sigma_8(0)$, which is used to normalize the power spectrum. Based
solely on fits to the linear growth rate at $z\lesssim 1$ \cite{Melia2017a}, $b\sigma_8(0)$
in $R_{\rm h}=ct$ is inferred to have the value $0.40\pm0.03$. The solid curves shown
in figure~2 are calculated with the normalizations $b\sigma_8(0)=0.37$ and $0.43$,
bracketing the $1\sigma$ variation of the best fit. The consistency between
the halo mass function at $z\sim 4-7$ and the linear growth rate at $z\lesssim 1$
\cite{Melia2017a} should not be underestimated. The fact that these two quite
different sets of measurements---the growth rate at $z\lesssim 1$ and the halo
mass function at $z\sim 4-7$---are mutually consistent with the expansion and
growth rates predicted by $R_{\rm h}=ct$ is a big factor in favor of this model. 

\begin{figure}
\begin{center}
\includegraphics[width=1.1\linewidth]{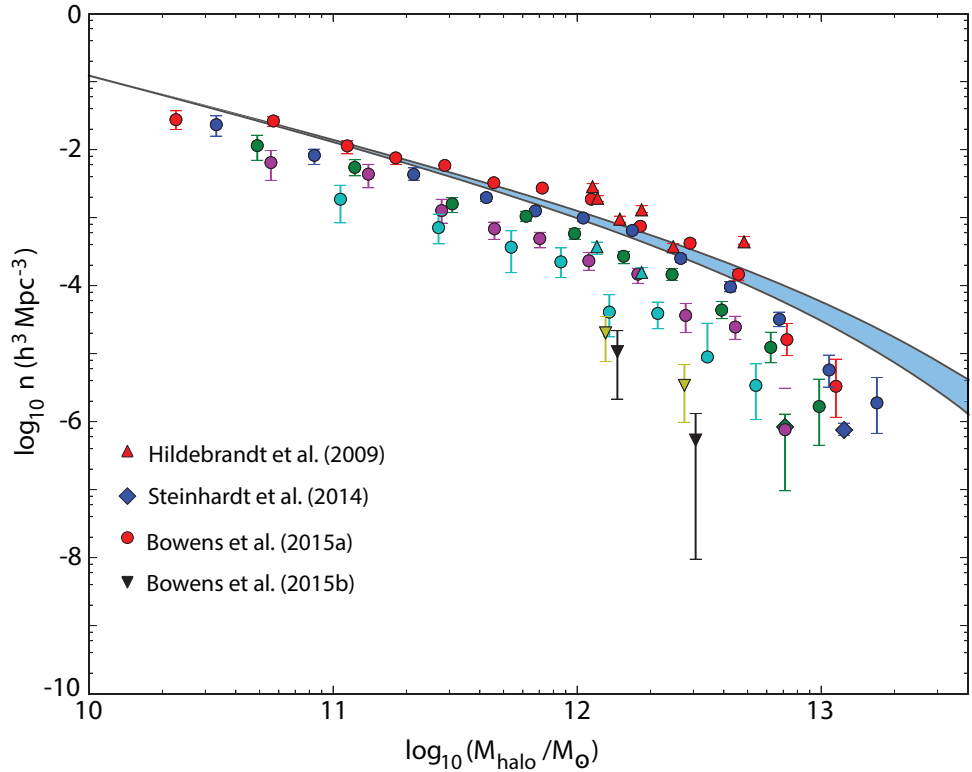}
\end{center}
\caption{Same as figure~1, except that the data have been recalibrated
for the $R_{\rm h}=ct$ universe using the ratio of differential comoving volumes
in Equation~(24). The solid curves represent the halo mass function for
$R_{\rm h}=ct$ calculated from Equations~(20) and (21), using a normalization
$b\sigma_8(0)=0.43$ and $0.37$. These two values bracket the $1\sigma$ variation
of the optimized fit to recently published redshift space distortion measurements
of the cosmological growth rate \cite{Melia2017a}, which produced the value
$b\sigma_8(0)=0.40\pm0.03$, with $b$ the growth factor in Eq.~(18). Given the
very weak dependence of $\delta_k(z)$ on redshift (Eq.~16), the halo mass function in this model
is essentially independent of $z$ in the range $4-10$.}
\end{figure}

Having said this, it is nonetheless also true that $R_{\rm h}=ct$ overpredicts 
the number of halos towards the high mass end, as one can see from the right-hand
side of figure~2. Several factors may be contributing to this: (1) As noted earlier,
the Bolshoi simulation indicates that the Sheth-Tormen mass function overpredicts 
the number of halos at high mass and high redshift, at least in the context of
$\Lambda$CDM. If an analogous result holds in $R_{\rm h}=ct$, this might be
the cleanest explanation for the theoretical overprediction seen in this figure.
To be certain of this effect, however, we must await the completion of a full
Bolshoi-like simulation with $R_{\rm h}=ct$ as the background cosmology. (2) 
The number of halos inferred at high mass and high redshift may be incomplete
due to observational selection effects and/or an incompleteness in the galaxy
surveys. In this scenario, the apparent overprediction of halos at high mass
and high redshift may get resolved with future observations. (3) As discussed
above, the equations we have used in this paper to explore the linear growth 
rate of fluctuations may break down at $z\gtrsim 10$ due to the increasing
role of radiation, which we have ignored until now. A more sophisticated 
analysis will be required at high redshifts to handle a more realistic
growth rate calculation.

By comparison, the predicted halo mass function in $\Lambda$CDM misses the data
everywhere on two counts: first, the shape of $n(M_{\rm halo})$ is not a good match 
to the observed distribution; $\Lambda$CDM predicts a steep function, in significant
tension with the measurements at all redshifts. Second, $\Lambda$CDM predicts a
strong evolution of the halo mass function with redshift, notably underpredicting
the observed density at high redshifts by many orders of magnitude, while at the
same time overpredicting their density at lower redshifts. Coupled to the standard
model's inability to properly account for the observed flat linear growth rate at
$z\lesssim 1$, our results in this paper suggest that the growth of structure in
the standard model is inconsistent with the data, at least out to $z\sim 10$.

There are several caveats to this conclusion, however. First, the steepness of the 
function predicted by $\Lambda$CDM is largely due to the assumption of a constant 
light to mass ratio in the conversion of luminosity to mass. The same assumption 
does not adversely affect $R_{\rm h}=ct$, but perhaps the tension with $\Lambda$CDM
may be alleviated (at least partially) with a more complicated relation between
light and mass that produces a shallower mass function. As we have discussed
previously, $\Lambda$CDM does not directly predict this function, relying
instead on theoretical adjustments to fit the data. It is more a question of
whether these adjustments produce other signatures that may or may not be in
tension with the observations. In other words, what see see in figure~1 is
essentially a rescaling of the galaxy luminosity function, not a direct
measurement of the halo mass function, and this rescaling relies on several
pieces of poorly understood physics. Eventually, it will be necessary to
carry out a comprehensive simulation that couples the cosmological
halo evolution with the physics of galaxy formation in order for us to
be more certain regarding the translation of galaxy data into a reliable
halo mass function. A second caveat is that even in $R_{\rm h}=ct$ the halo 
mass function deviates significantly from the predicted value for $z>7$, 
as we have discussed above. The fact that both models show significant
deviations at the high mass, high redshift end of the halo distribution
may therefore be an indication that at least some of the tension with 
the data is related to the poorly known underlying physics, rather than
it merely being an indication that the issue lies solely with the
cosmology.

\section{Conclusion}
Some evidence suggests a disparity between the halo mass function
predicted by $\Lambda$CDM and the actual measurements of this quantity.
This work has been motivated by its lack of anticipated strong evolution
in redshift and a significant steepening of the distribution with increasing
mass. The observed halo mass function is relatively flat, pointing to a much
higher density at the high-mass end, and a correspondingly lower density of
halos with smaller masses. In addition, the halo distribution appears to be
evolving much more slowly than is required by the standard model, at least
for redshifts $z\lesssim 10$.

In earlier work, we used recently published redshift space distortion
measurements of the cosmological growth rate, $b\sigma_8(z)$, at
redshifts $z\lesssim 1$, to show that the linear evolution of perturbations
in the $R_{\rm h}=ct$ cosmology appears to be a better match to the data
than the current standard model. In that work, we found an optimized
growth rate $b\sigma_8(0)=0.40\pm0.03$. Interestingly,
the halo mass function in $R_{\rm h}=ct$, normalized using this value,
fits the measured halo distribution very well, certainly much better
than $\Lambda$CDM does. Of course, one must still be wary of these
results, given that the halo masses cannot be measured directly.
Nonetheless, since the various techniques used to translate the
observed galaxy mass distribution into a halo mass function all
agree qualitatively on the outcome, the mutual consistency between
the halo fits at $z\sim 4-10$ and the linear growth rate fit at
$z\lesssim 1$, suggests that the formation of structure predicted
by $R_{\rm h}=ct$ is a better fit to the observations than
that expected in $\Lambda$CDM.

Although our focus in this paper has been exclusively on the
formation and evolution of halos, the surprising nature of the
halo mass function has also been characterized as a breakdown in
the theory of galaxy formation at high redshifts. But this conclusion
is actually not new. It has been known for several
years that the observed high-$z$ quasars and galaxies must have
formed much too quickly when viewed with the timeline afforded
them by the standard model \cite{Melia2013,Melia2014,Melia2015}.
For example, in order to understand the emergence of $\sim 10^9\;
M_\odot$ black holes earlier than $z\sim 6-7$ in $\Lambda$CDM, one must
assume that black-hole growth either began with anomalously large seeds
($M> 10^5\;M_\odot$), or proceeded at super-Eddington rates, neither of
which has ever been seen---here locally, or at high redshifts. Yet
in $R_{\rm h}=ct$, the timeline is just right to have allowed these
high-$z$ objects to form according to well known and understood
astrophysical principles.

Our analysis of the halo mass function in this paper
constitutes another important confirmation of the expansion scenario
predicted by the $R_{\rm h}=ct$ cosmology. But there is much work
yet to be done with the formation of structure in this model. Though
$R_{\rm h}=ct$ accounts for the observed halo distribution very well
for masses $M\lesssim 10^{12}\;M_\odot$, its predictions deviate
from the observations by one to two orders of magnitude at the highest
mass end. Though this is still much less extreme than the situation
one confronts in $\Lambda$CDM, this divergence suggests several
possible explanations that need to be explored. These include (1) the
linear growth rate seen at low redshifts in $R_{\rm h}=ct$ may need
modification at $z\gtrsim 10$. This would not be surprising, given that
the current theory is based on the dominance of matter and dark energy
in the cosmic fluid. But at high redshifts, the growing relevance of
radiation cannot be ignored; and (2) The relations used to infer the
halo masses from large galaxy surveys ought to be revisited. Attempts
at mitigating the disparity between theory and observation in
$\Lambda$CDM had very little impact. They may be more successful
in the case of $R_{\rm h}=ct$, since the current tension exists
solely for $M\gtrsim 10^{12}\;M_\odot$. The halo mass function for
masses lower than this agrees with theory very well.

Eventually, a complete study of galaxy formation and clustering
in $R_{\rm h}=ct$ needs to be carried out in concordance with the
evolution of the halo mass function.
Hopefully, this self-consistent approach will mitigate at least several
of the uncertainties still hindering the interpretation of halo
properties from the actual measurement of the galaxy luminosity
function and spectrum. Currently, there is still too much freedom in
combining various physical influences on galaxy evolution to arrive
at a unique picture of structure formation at redshifts $z\lesssim 10$.

\section*{Acknowledgments}
We are very grateful to the anonymous referee for suggesting
several important improvements to the presentation of our results
in this paper. FM is grateful to the Instituto de Astrof\'isica
de Canarias in Tenerife and to Purple Mountain Observatory in Nanjing, China
for their hospitality while part of this research was carried out. FM is also
grateful for partial support to the Chinese Academy of Sciences Visiting
Professorships for Senior International Scientists under grant 2012T1J0011,
and to the Chinese State Administration of Foreign Experts Affairs under
grant GDJ20120491013.


\begin{thebibliography}{90}
\bibitem[1]{Sheth2001} R. K. Sheth, H. J. Mo and G. Tormen, MNRAS {\bf 323} 1 (2001) 1.
\bibitem[2]{Springel2005} V. Springel, S.D.M. White, A. Jenkins et al., Nature {\bf 435} (2005) 629.
\bibitem[3]{Vogelsberg2014} M. Vogelsberger, S. Genel, V. Springel et al., MNRAS {\bf 444} (2014) 1518.
\bibitem[4]{Press1974} W. Press and P. Schechter, ApJ {\bf 187} (1974) 425.
\bibitem[5]{Steinhardt2016} C. L. Steinhardt, P. Capak, D. Masters and J. S. Speagle, ApJ {\bf 824} (2016) 1.
\bibitem[6]{Grogin2011} N. A. Grogin, D. D. Kocevski, S. M. Faber et al., ApJS {\bf 197} (2011) 35.
\bibitem[7]{Koekemoer2011} A. M. Koekemoer, S. Faber, H. Ferguson et al., ApJS {\bf 197} (2011) 36.
\bibitem[8]{Capak2016} P. Capak et al., in preparation (2016).
\bibitem[9]{Hildebrandt2009} H. Hildebrandt, J. Pielorz, T. Erben et al., A\&A {\bf 498} (2009) 725.
\bibitem[10]{Lee2012} K.-S. Lee, H. C. Ferguson, T. Wiklind et al., ApJ {\bf 752} (2012) 66.
\bibitem[11]{Ilbert2013} O. Ilbert, H. J. McCracken, O. Le F\'evre et al., A\&A {\bf 556} (2013) A55.
\bibitem[12]{Finkelstein2015} S. L. Finkelstein, M. Song, P. Behroozi et al., ApJ {\bf 814} (2015) 95.
\bibitem[13]{Caputi2015} K. I. Caputi, O. Ilbert, C. Laigle et al., ApJ {\bf 810} (2015) 73.
\bibitem[14]{Lacey2011} C. G. Lacey, C. M. Baugh, C. S. Frenk, and A. J. Benson, MNRAS {\bf 412} (2011) 1828.
\bibitem[15]{Finlator2011} K. Finlator, B. D. Oppenheimer, and R. Dav\'e, MNRAS {\bf 410} (2011) 1703.
\bibitem[16]{Waters2016} D. Waters, S. M. Wilkins, T. Di Matteo, Y. Feng, R. Croft and D. Nagai, MNRAS {\bf 461} (2016) L51.
\bibitem[17]{Behroozi2015} P. S. Behroozi and J. Silk, ApJ {\bf 799} (2015) id 32.
\bibitem[18]{Mashian2016} N. Mashian, P. A. Oesch and A. Loeb, MNRAS {\bf 455} (2016) 2101.
\bibitem[19]{Bouwens2015a} R. J. Bouwens, G. D. Illingworth, P. A. Oesch et al., ApJ {\bf 803} (2015) 34.
\bibitem[20]{Bouwens2015b} R. J. Bouwens, P. A. Oesch, I. Labbe et al., ApJ {\bf 830} (2015) id 67.
\bibitem[21]{Melia2013} F. Melia, ApJ {\bf 764} (2013) 72.
\bibitem[22]{Melia2015} F. Melia and T. M. McClintock, Proc. R. Soc. A {\bf 471} (2015) 20150449.
\bibitem[23]{Melia2014} F. Melia, AJ {\bf 147} (2014) 120.
\bibitem[24]{Leauthaud2012} A. Leauthaud, J. Tinker, K. Bundy et al., ApJ {\bf 744} (2012) 159.
\bibitem[25]{Planck2016} Planck Collaboration, P.A.R. Ade, N. Aghanim et al., A\&A {\bf 594} (2016) id A13.
\bibitem[26]{Melia2017a} F. Melia, MNRAS {\bf 464} (2017) 1966.
\bibitem[27]{Melia2007} F. Melia, MNRAS {\bf 382} (2007) 1917.
\bibitem[28]{Melia2016a} F. Melia, Frontiers of Phys. {\bf 11} (2016) 118901.
\bibitem[29]{Melia2017b} F. Melia, Frontiers of Phys. {\bf 12} (2017) 129802.
\bibitem[30]{Melia2009} F. Melia and M. Abdelqader, IJMP-D {\bf 18} (2009) 1889.
\bibitem[31]{Melia2012} F. Melia and A. Shevchuk, MNRAS {\bf 419} (2012) 2579.
\bibitem[32]{Murray2013} S. G. Murray, C. Power and A.S.G. Robotham, A\&C {\bf 3} (2013) 23.
\bibitem[33]{Conroy2009} C. Conroy, J. E. Gunn, and M. White, ApJ {\bf 699} (2009) 486.
\bibitem[34]{Schaerer2010} D. Schaerer and S. de Barros, A\&A {\bf 515} (2010) A73.
\bibitem[35]{Stark2013} D. P. Stark, M. A. Schenker, R. Ellis, B. Robertson, R. McLure and J. Dunlop, ApJ {\bf 763} (2013) 129.
\bibitem[36]{Speagle2014} J. S. Speagle, C. L. Steinhardt, P. L. Capak and J. D. Silverman, ApJ {\bf 214} (2014) 15.
\bibitem[37]{Ma2017} Ma Xiangcheng, Philip F. Hopkins, Shea Garrison-Kimmel et al., arxiv:170606605M (2017)
\bibitem[38]{Stefanon2017} Mauro Stefanon, Rychard J. Bouwens, Ivo Labb\'e, ApJ {\bf 843} (2017) 36S.
\bibitem[39]{Ceverino2017} Daniel Ceverino, Simon C.O.Glover, Ralf S. Klessen, MNRAS {\bf 470} (2017) 2791C.
\bibitem[40]{O'Shea2015} Brian W.Oshea, John H. Wise, Hao Xu, et al., ApJ {\bf 807} (2015) 12O. 


\bibitem[41]{Trac2015} Trac H., Cen R., Mansfield P., 2015, ApJ, 813, 54T.
\bibitem[42]{Weinberg1972} S. Weinberg, Gravitation and Cosmology, Wiley: NY (1972).
\bibitem[43]{Landau1975} L. D. Landau and E. M. Lifshitz, The Classical Theory of Fields,
Pergamon Press: Oxford (1975).
\bibitem[44]{Peebles1980} P.J.E. Peebles, The Large-scale Structure of the Universe,
Princeton University Press: NY (1980).
\bibitem[45]{Peebles1993} P.J.E. Peebles, Principles of Physical Cosmology, Princeton University Press: NY (1993).
\bibitem[46]{Press1980} W. H. Press and E. T. Vishniac, ApJ {\bf 239} (1980) 1.
\bibitem[47]{Kolb1990} E. W. Kolb and M. Turner, The Early universe, Addison-Wesley: Reading MA (1990).
\bibitem[48]{Padmanabhan1993} T. Padmanabhan, Structure Formation in the Universe,
Cambridge University Press: Cambridge, England (1993).
\bibitem[49]{Coles1995} P. Coles and F. Lucchin, Cosmology: The Origin and Evolution
of Cosmic Structure, Wiley: NY (1995).
\bibitem[50]{Peacock1999}J. A. Peacock, Cosmological Physics, Cambridge University
Press: Cambridge, England (1999).
\bibitem[51]{Liddle2000} A.R.D. Liddle and K. A. Lyth, Cosmological Inflation
and Large-Scale Structure, Cambridge University Press: Cambridge, England (2000).
\bibitem[52]{Tsagas2008} C. G. Tsagas, A. Challinor and R. Maartens, Phys. Reports {\bf 465} (2008) 61.
\bibitem[53]{Klypin2011} Klypin, A. A., Trujillo-Gomez, S. \& Primack, J., ApJ {\bf 740} (2011) 102.
\end{thebibliography}
\end{document}